\documentclass[10pt,journal,final,twocolumn]{IEEEtran}

\usepackage{graphics}
\usepackage{subfigure}
\usepackage{epstopdf}
\usepackage{epsfig}
\usepackage[cmex10]{amsmath}
\usepackage{amssymb}
\usepackage{times}
\usepackage{ntheorem}
\usepackage{pifont}
\usepackage{stfloats}
\usepackage{bm}
\usepackage{color}
\usepackage{makecell}
\usepackage{array}
\usepackage{multirow}

\title{Sensing-Aided Channel Estimation for Near-Field MIMO ISAC Systems via Cross-Attention Transformer}

\author{Peihao Dong,~\IEEEmembership{Member,~IEEE}, Renbin Li, Shen Gao, Shuangshuang Li, Fuhui Zhou,~\IEEEmembership{Senior Member,~IEEE},\\
Wei Xu,~\IEEEmembership{Fellow,~IEEE}, and Qihui Wu,~\IEEEmembership{Fellow,~IEEE}
\thanks{

P. Dong, R. Li, S. Li, and Q. Wu are with the College of Electronic and Information Engineering, Nanjing University of Aeronautics and Astronautics, Nanjing 211106, China (e-mail: phdong@nuaa.edu.cn; renbinli@nuaa.edu.cn; lishsh@nuaa.edu.cn; wuqihui2014@sina.com).
	
S. Gao is with the College of Telecommunications and Information Engineering, Nanjing University of Posts and Telecommunications, Nanjing 210003, China, and also with the National Mobile Communications Research Laboratory, Southeast University, Nanjing 211111, China (e-mail: gaoshen@njupt.edu.cn).
	
F. Zhou is with the College of Artifcial Intelligence, Nanjing University of Aeronautics and Astronautics, Nanjing 211106, China (e-mail: zhoufuhui @ieee.org).
	
W. Xu is with the National Mobile Communications Research Laboratory, Southeast University, Nanjing 211111, China (e-mail: wxu@seu.edu.cn).}%
}

\begin{document}
	\maketitle
\begin{abstract}
Near-field integrated sensing and communication (ISAC) can deliver the high spatial resolution and transmission capability with the shared spectrum and hardware. Due to the partial overlap between communication scatterers and radar targets, the sensing information can provide valuable priors to enhance the channel estimation while fusing the two heterogeneous modalities remain challenging. To address this problem, a Cross-Attention Transformer based Channel Estimation Neural Network (CAT-CENet) is developed, which includes a communication pilot branch generating the the Key and Value features and a sensing information branch generating the Query feature. By elaborating the three-module structure, CAT-CENet can focus on  features of overlapped targets automatically without need of identifying them in advance. The modality contribution is theoretically analyzed based on the Shapley value to verify the cross-attention gain achieved by CAT-CENet. Simulation results show that CAT-CENet outperforms the state-of-the-art schemes, especially with the higher overlapping proportion, and is robust to the model pruning.
\end{abstract}


\section{Introduction}
The emerging demands of massive connectivity and high-precision environment sensing have intensified challenges of spectrum scarcity and network scalability, which can be potentially treated by extremely large-scale MIMO (XL-MIMO) and integrated sensing and communication (ISAC). The XL-MIMO significantly improves the spatial-resolution and multiplexing capability by enlarging the array size \cite{H. Lu}, \cite{Z. Wang}, \cite{W. Hu}. The ISAC system facilitates sharing the spectrum and hardware via the joint design of communication and sensing functions \cite{J. Li}, \cite{F. Liu}. The integration of XL-MIMO and ISAC is thus seen as a promising direction for future wireless systems.

In near-field XL-MIMO scenarios, the assumptions of far-field plane-wave propagation no longer hold due to the large physical aperture of antenna arrays. The resulting spherical wavefront leads to distance-dependent channel characteristics and significantly increases channel dimensionality \cite{X. Mu}. To address the complexity of such channels, prior studies have explored compressed sensing techniques, including orthogonal matching pursuit \cite{M. Cui} and bayesian prior estimation \cite{J. Cao}. These methods depend on stringent sparsity assumptions and typically entail high computational costs. Meanwhile, deep learning (DL) based approaches have been applied to XL-MIMO channel estimation to reduce complexity and improve performance, but robustness under high-dimensional noisy channels is still limited \cite{W. Xu, S. Gao, B. Zhou, S. Li}. In the context of ISAC channel estimation, a tensor-based approach was proposed in \cite{R. Zhang} to jointly estimate channel state information and target parameters. In \cite{S. Xu}, a Brownian Bridge-based diffusion channel denoising method was applied to massive MIMO ISAC systems. In \cite{Y. Liu}, a three-stage deep learning framework was proposed for intelligent reflecting surface assisted ISAC systems to estimate direct and reflected channels.

Since communication scatterers and radar targets often partially overlap \cite{Z. Huang}, the communication channel estimation can be enhanced by using the radar sensing information. However, the inherent modality heterogeneity between the communication and sensing information makes the fusion task intractable and the dependence on the accurate overlapping knowledge exacerbates the challenge. Therefore, a cross attention transformer-based channel estimation neural network (CAT-CENet) is proposed in this article. The main novelty and contribution are summarized as follows.
	
1) To better accommodate the modality heterogeneity, the cross-attention is adopted to fuse the communication and sensing information instead of the widely used self-attention. The overlapped targets need not to be identified in advance since CAT-CENet can focus on their features automatically via the end-to-end learning.

2) CAT-CENet is composed of three modules, which have respective functions and collaborate with each other. The preprocessing module include two branches respectively in charge of generating the Query feature and generating the Key and Value features that can be better grasped via the convolutional embedding operation. The cross-modality fusion module includes the serial cross-attention and dual-attention blocks for each encoder, where the former helps the network focus on features of overlapped targets based on the calculated cross-attention matrix and the latter further fuses the sensing information into the pilot feature from different dimensions of the feature tensor. The post-processing module connects the previous two modules via the residual structure to output the purified channel matrix.

3) A performance metric, Shapley value, is first applied to theoretically quantify the contributions of input tensors corresponding to the communication pilot and sensing information of different targets, which helps visualize the feature contribution change under different overlapping conditions and thus verify the cross-modality gain. Simulation comparison with the state-of-the-art schemes shows the superiority of the proposed CAT-CENet, which can be enlarged with the higher overlapping proportion. The pruned CAT-CENet can maintain the performance with the reduced model size.

\section{System Model}


\begin{figure}[t!]
	\centering
	\includegraphics[width=0.47\textwidth]{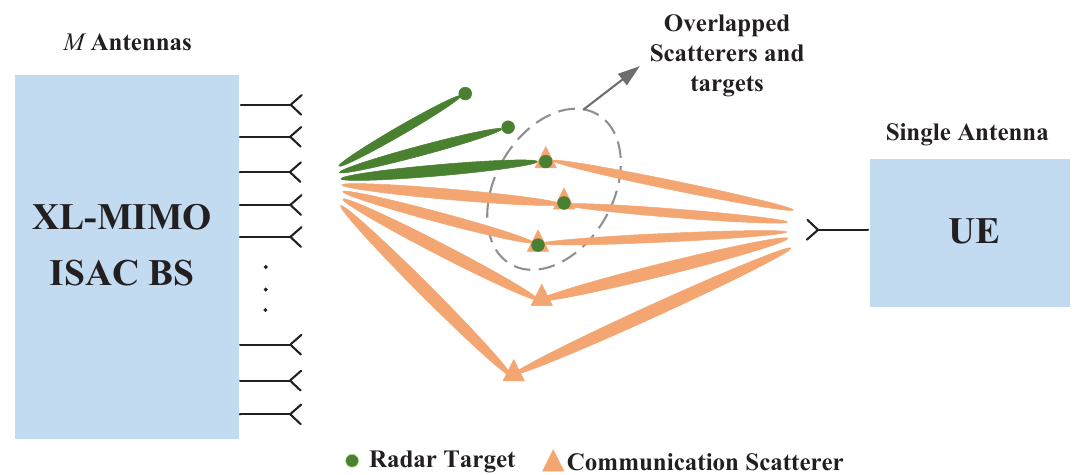}
	\caption{The XL-MIMO ISAC system with overlapped scatterers and targets.}
	\label{fig}
\end{figure}

We consider a near-field time-division duplexing (TDD) XL-MIMO ISAC system model illustrated in Fig. 1, where a base station (BS) with $M$ antennas serves a single-antenna mobile user via an $L$-path channel while detects $K$ radar targets. With the radar-only sensing mode, the BS transmits probing signals and receives the echoes reflected by surrounding targets to estimate their positions. Some radar targets also serve as communication scatterers of the BS-user channel \cite{C. Masouros}.
Denote the communication scatterer set as $\mathcal{C} = \{C_{1}, C_{2}, \ldots, C_{L}\}$ and the sensing target set as $\mathcal{S} = \{S_{1}, S_{2}, \ldots, S_{K}\}$. Then their overlapping relationship can be denoted as $\mathcal{D} = \mathcal{C} \cap \mathcal{S} = \{D_{1}, D_{2}, \ldots, D_{X}\}$, where the set cardinality satisfies $\quad 1 \leq X \leq \min(K,L)$. Due to the partial overlap, the radar sensing information is useful to enhance the communication channel estimation.

To estimate the channel vector between the user and the BS $ \mathbf{h}_\mathrm{c} \in \mathbb{C}^{M \times 1}$, the user transmits the pilot $\sqrt{P}x$ with the power $P$. Then the received pilot signal at the BS $\mathbf y$ is given by
\begin{equation}
\label{eqn_y}
\mathbf{y} = \sqrt{P}\mathbf{h}_\mathrm{c} x +\mathbf{n},
\end{equation}
where $ \mathbf{n} \sim \mathcal{C} \mathcal{N} \left(0, \sigma^2 \mathbf{I} \right)$ represents the additive white Gaussian noise (AWGN) with variance $ \sigma^2$. In an XL-MIMO ISAC system, the separation of near-field and far-field propagation effects is determined by the Rayleigh distance. Assuming a uniform linear array (ULA) with $M$ antennas and the adjacent antenna spacing $d=\frac{\lambda}2$, the Rayleigh distance is given by $D_{\text{Ray}}= \frac{2 D_\mathrm{a}^2}{\lambda} = \frac{1}{2} M^2 \lambda $ with ${\lambda}$ the carrier wavelength, indicating the near-field effect should be considered when the XL-MIMO is used at the BS. Then the near-field communication channel is modeled as
\begin{equation}
\mathbf{h}_\mathrm{c} = \sqrt{\frac{M}{L}} \sum_{l=1}^{L} g_l \mathbf{a}(\phi_l, r_l)\label{eq},
\end{equation}
where $ g_l$ and $\mathbf{a}(\phi_l, \!r_l)\! =\! \frac{1}{\sqrt{M}}\! [ e^{-\!j \frac{2\pi}{\lambda} (r_{l,\!1}\! -\! r_l)}, \!\ldots\!, e^{-\!j \frac{2\pi}{\lambda} (r_{l,\!M} \!-\! r_l)} ]^T$ respectively denote the path gain and the array response vector at the BS with the azimuth angle of arrival (AoA) $\phi_l \in \left[ -\frac{\pi}{2}, \frac{\pi}{2} \right]$ and the distance $r_l $ for the $l$-th path. Specifically, $r_l $ and $ r_{l,m} = \sqrt{r_l^2 + \delta_m^2 d^2 - 2 r_l \delta_m d \sin \phi_l}$ represent the distance from the $l$-th scatterer to the center and the $m$th antenna of the BS array with $\delta_m = \frac{2m - M - 1}{2}, m = 1, \ldots, M$.
	
The radar sensing is actually a round-trip process of the probing signal, which is transmitted by the BS and then reflected by the targets back to the BS. Thus the radar sensing channel matrix can be modeled as
\begin{equation}
\mathbf{H}_\mathrm{s} = \sum_{k=1}^{K} s_k \mathbf{a}(\theta_k, v_k) \mathbf{a}^\mathrm{H}(\theta_k, v_k),
\end{equation}
where $s_k$ and $\mathbf{a}(\theta_k, v_k) \in \mathbb{C}^{M \times 1}$ respectively denote the radar cross section (RCS) of the $k$th target and the array response vector at the BS with the azimuth AoA $\theta_k\in\left[ -\frac{\pi}{2}, \frac{\pi}{2} \right]$ and the distance between the $k$th target and the center of the BS array $v_k $.
	
In contrast to the conventional far-field channel mainly depending on the angular information, the near-field channel is determined by not only the path angle but also the path distance and thus has a quite different sparsity domain, which compels traditional methods to identify the far or near field in advance. Though DL-based methods can get rid of this requirement, the network still needs to be elaborated to compensate the performance loss brought by the additional distance information as well as to overcome the challenge of the heterogeneous modality fusion.

\begin{figure*}[t]
	\centering
	\includegraphics[width=6.3in]{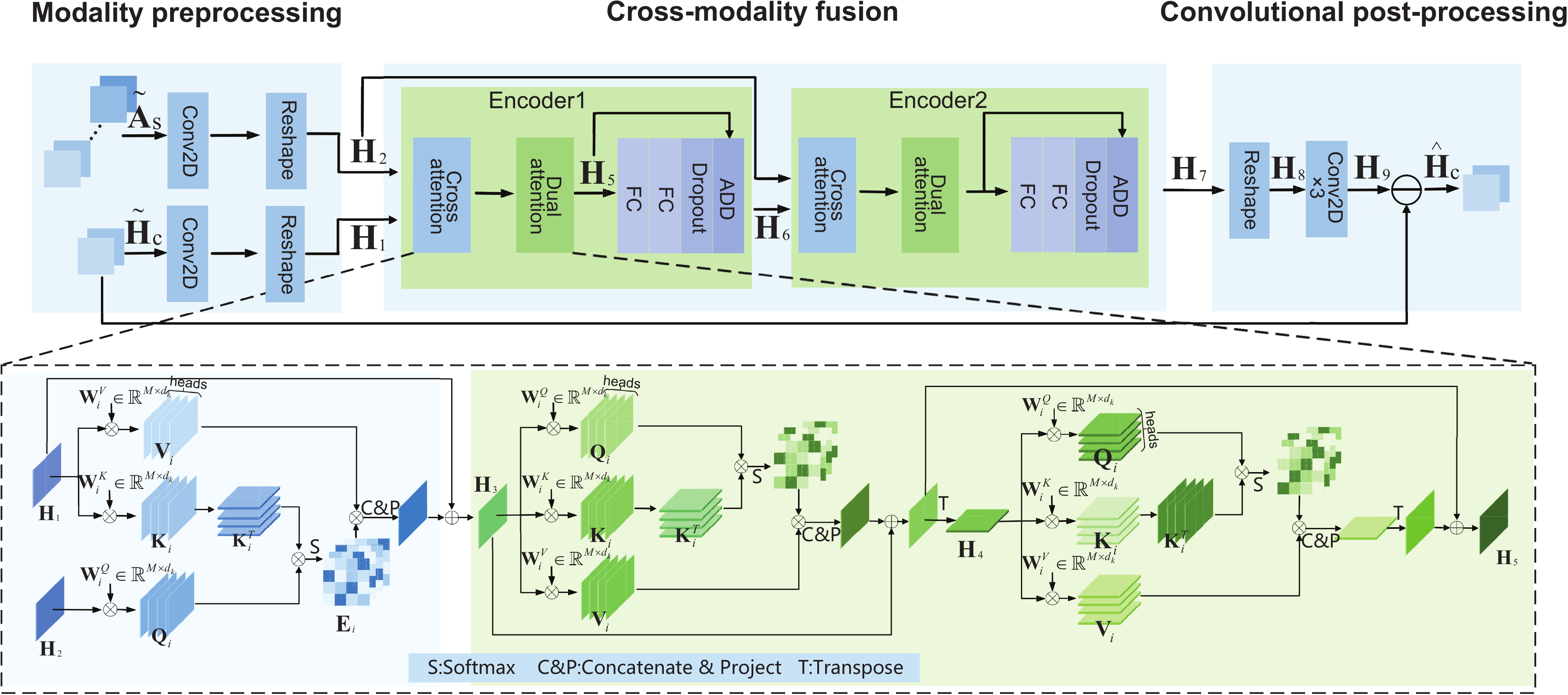}  
    \caption{Illustration of the overall structure of CAT-CENet and Cross-Modality Attention mechanism of Encoder.}
	\label{fig:wide-image}
\end{figure*}

\section{Proposed CAT-CENet}

In this section, CAT-CENet centered on the cross-modality fusion is proposed to enhance the communication channel estimation by exploiting the radar sensing information. The network with the three-module structure is first elaborated, after which the complexity is analyzed. Finally, the modality contribution is analyzed based the Shapley value to helps verify the cross-modality gain.

\subsection{Network Inputs}\label{AA}
	
By letting $x=1$ in (\ref{eqn_y}) without loss of generality, the least square (LS) estimate of $\mathbf{h}_\mathrm{c}$ is expressed by
\begin{equation}
\tilde{\mathbf{h}}_\mathrm{c} = \frac{\mathbf{y}}{\sqrt{P}} = \mathbf{h}_\mathrm{c} + \frac{\mathbf{n}\label{eq}}{\sqrt{P}},
\end{equation}	
which will be refined by CAT-CENet to yield a more accurate version. By splitting the real and imaginary parts, $\tilde{\mathbf{h}}_\mathrm{c}$ and ${\mathbf{h}_\mathrm{c}}$ are  reshaped into tensors $\tilde{\mathbf{H}}_\mathrm{c} \in \mathbb{R}^{\sqrt{M} \times \sqrt{M} \times 2}$ and $\mathbf{H}_\mathrm{c} \in \mathbb{R}^{\sqrt{M} \times \sqrt{M} \times 2}$ to act as the communication input and the label, respectively.

In the ISAC system, the communication channel is estimated after the initial sensing of targets, also known as target searching, which can be carried out via array signal processing techniques \cite{C. Masouros}. In this article, we dedicate to enhancing the communication channel estimation by exploiting the pre-estimated target information with estimation errors incorporated. Specifically, the estimated AoA and distance of $K$ targets are denoted as $\hat{\theta}_k=\theta_k-\epsilon_{\theta_k^r}$ and $\hat{v}_k=v_k-\epsilon_{v_k}$, respectively, with $\epsilon_{\theta_k^r}$ and $\epsilon_{v_k}$ the corresponding errors for $k=1,...,K$. By splitting the real and imaginary parts, the estimated response vectors $\mathbf a (\hat{\theta}_k,\hat{v}_k)$ are reshaped into the tensor $\tilde{\mathbf{A}}_k\in \mathbb{R}^{\sqrt{M} \times \sqrt{M} \times 2}$ for $k=1,...,K$. By combining these $K$ tensors, the input tensor of the radar sensing branch, $\tilde{\mathbf{A}}_{\mathrm{s}}\in \mathbb{R}^{\sqrt{M} \times \sqrt{M}\times 2K}$, is obtained. It is noted that the number of targets, $K$, can be obtained from the preceding target tracking step and thus is not the additional prior knowledge. The number of scatterers, $L$, is unknown since the pilot matrix is processed as a whole. The number of overlapped targets, $X$, also needs not to be known since CAT-CENet can identify their features automatically, making CAT-CENet get rid of the dependence on the overlapping knowledge.

\subsection{Network Structure}

As shown in Fig. 2, CAT-CENet is composed of three modules, which have respective specific functions and collaborate with each other. The preprocessing module coverts input tensors into more tractable feature maps via the convolutional embedding operation. The cross-modality fusion module plays vital roles in finding and exploiting features corresponding to overlapped targets to enhance the feature extraction of the communication channel. The covolutional post-processing module extracts the noise and outputs the estimated communication channel via the cross-module residual connection.

\subsubsection{Modality Preprocessing}
The input tensors $\tilde{\mathbf{H}}_{\mathrm{c}}$ and $\tilde{\mathbf{A}}_{\mathrm{s}}$ are respectively processed by the two-dimensional (2D) convolutional layer with $F$ $3 \times 3$ kernels, yielding two $\sqrt{M} \times \sqrt{M} \times F$ feature maps. This step acts as the embedding operation transforming input tensors into more tractable features. By combing the first two dimensions, the two feature maps are reshaped into $\mathbf{H}_1 \in \mathbb{R}^{M \times F}$ and $\mathbf{H}_2 \in \mathbb{R}^{M \times F}$, respectively. 

\subsubsection{Cross-modality Fusion}
	
To fully fuse the heterogeneous features $\mathbf{H}_1$ and $\mathbf{H}_2$, this module includes two encoders, each of which has a cross-attention block and a dual-attention block. The cross-attention block aims to learn structural characteristics of the communication channel, aided by the sensing information. The dual-attention block including the feature map attention and the spatial attention is appended to further fuse the sensing information into the pilot feature from different dimensions of the feature tensor for the efficient noise extraction.

Encoder1 starts with the generation of Query, Key and Value matrices. Instead of using the communication and sensing inputs together to calculate each of them, CAT-CENet allocates the generation of Query matrix to the sensing branch by using $\mathbf{H}_2$ and the generation of Key and Value matrices to the communication branch by using $\mathbf{H}_1$ to facilitate the calculation of the cross-attention matrix. To this end, the multi-head cross-attention is adopted with the number of heads $J$. For the $j$th head, we have
\begin{equation}
	\mathbf{Q}_{ j} = \mathbf{H}_2\mathbf{ W}_{ j}^\mathrm{Q},  \mathbf{K}_{ j} = \mathbf{H}_1\mathbf{ W}_{ j}^\mathrm{K},  \mathbf{V}_{ j} = \mathbf{H}_1\mathbf{ W}_{ j}^\mathrm{V},
\end{equation}
where $\mathbf{W}_{j}^\mathrm{Q} \in \mathbb{R}^{F \times F}$, $\mathbf{W}_{j}^\mathrm{K} \in \mathbb{R}^{F \times F}$, and $\mathbf{W}_{j}^\mathrm{V} \in \mathbb{R}^{F \times F}$ denote the corresponding learnable projection matrices. Then the attention matrix of the $j$th head is calculated as
\begin{equation}
\mathbf{ E}_{ j} = \text{Softmax}\left( \frac{\mathbf{Q}_{j} \mathbf{K}_{ j}^\mathrm{T}}{\sqrt{F}} \right),	
\end{equation}
which used to capture the common structural feature between the communication and sensing information corresponding to overlapped scatterers. The output of the the $j$th head is given by
\begin{equation}
\mathbf{Z}_j =\mathbf{ E}_{ j}\mathbf{ V}_j ,	
\end{equation}
which enhances the feature extraction of overlapped scatterers in the communication information $\mathbf{ V}_i$ guided by the attention matrix $\mathbf{ E}_{ j}$. By concatenating the outputs of all heads, a linear transformation matrix $\mathbf{W}^\mathrm{O}\in \mathbb{R}^{JF \times F}$ is applied to project the output features back to a compact feature with the original size of $M \times F$, that is
\begin{equation}
\mathbf{ Z} =  \text{Concat}(\mathbf{ Z}_1, \mathbf{ Z}_2, \dots, \mathbf{ Z}_h) \mathbf{W}^\mathrm{O}.
\end{equation}
The resulting $\mathbf{ Z}$ is then added to one of the original inputs $\mathbf{ H}_1$ via a shortcut connection to form $\mathbf{ H}_3\in \mathbb{R}^{{M} \times F}$, which improves training stability and mitigates issues such as gradient explosion.

The output of the cross-attention module $\mathbf{ H}_3$ is further refined through a dual-attention block across both the feature map and spatial dimensions, which effectively extracts the noise component and thereby improves the accuracy and robustness of channel estimation. In the first feature map attention sub-block, the multi-head self-attention is applied to generate Query, Key and Value matrices based on $\mathbf{ H}_3$. After that, the operations are similar to those of the cross-attention block mentioned-above, yielding the output of this sub-block $\mathbf{H}_4^{'} \in \mathbb{R}^{M \times F}$. Then $\mathbf{H}_4^{'}$ is transposed to $\mathbf{H}_4 \in \mathbb{R}^{F \times M}$ to align with the dimension of the following spatial attention sub-block. The processing is similar to the feature map attention sub-block. The feature map after the concatenation and projection is transposed to $\mathbf{H}_5^{'} \in \mathbb{R}^{M \times F}$, in order to add with $\mathbf{H}_4^{'}$ to obtain the output $\mathbf{ H}_5$. After a series of regular operations including two fully-connected (FC) layers, dropout and addition on $\mathbf{ H}_5$, the output of Encoder1 $\mathbf{ H}_6 \in \mathbb{R}^{M \times F}$ is obtained. Encoder2 repeats the operations of Encoder1 by using $\mathbf{ H}_6$ as the communication input and $\mathbf{ H}_2$ as the sensing input. This is because the communication pilot information acts the dominating role that needs to be processed constantly for the noise extraction while the sensing information only acts the auxiliary role during the processing. The output feature of Encoder2 is denoted as $\mathbf{ H}_7 \in \mathbb{R}^{M \times F}$.

\subsubsection{Convolutional Post-processing}
The feature map $\mathbf{H}_7$ is first reshaped into $\mathbf{H}_8 \in \mathbb{R}^{\sqrt{M} \times \sqrt{M} \times F}$ and then processed by three successive convolutional layers to obtain $\mathbf{H}_9 \in \mathbb{R}^{\sqrt{M} \times \sqrt{M} \times 2}$. Specifically, the first two convolutional layers with $64$ $3\times 3$ kernels are designed to extract the noise component embedded in the channel observations, while the last convolutional layer $2$ $3\times 3$ kernels generates the noise tensor $\mathbf{H}_9$. Finally, a cross-module residual connection is applied to subtract the extracted noise from the noisy channel $\tilde{\mathbf{H}}_{\mathrm{c}}$, obtaining the final channel estimate $\hat{\mathbf{H}}_\mathrm{c}$. Unlike the regular channel estimation networks, CAT-CENet benefits from both the residual denoising and auxiliary sensing information via the cross-modality residual backbone. CAT-CENet is trained to minimize the mean squared error loss expressed as
\begin{equation}
\mathcal{L} = \frac{1}{N_{\mathrm{tr}}} \sum_{i=1}^{N_{\mathrm{tr}}} \left\| {{\mathbf{H}}}^{(i)}_\mathrm{c} - \hat{\mathbf{H}}^{(j)}_\mathrm{c} \right\|_F^2,
\end{equation}
where $N_{\text{tr}}$ denotes the number of training samples and the superscript ($i$) indicates the $i$th sample with $\|\cdot\|_F^2$ Frobenius norm.

\subsection{Complexity Analysis}
This subsection analyzes the inference computational complexity of CAT-CENet in terms of floating-point operations (FLOPs), which includes five convolutional layers and two encoders. The overall complexity is expressed as
\begin{align}
	C_{\text{CAT-CENet}} \sim \mathcal{O}\bigg(
	& \sum_{i=1}^{5} N_\mathrm{x} N_\mathrm{y} G^2 C_{i,\text{in}} C_{i,\text{out}}  + 4(5JMF^2   \notag \\
	& + 4JM^2F + Q_{\text{in}} Q_{\text{out}})\bigg),\notag
\end{align}
where $N_\mathrm{x}$ and $N_\mathrm{y}$ are the length and width of the output feature map, $G$ is the length of side of the kernel, $C_{i,\text{in}}$ and $C_{i,\text{out}}$ are numbers of the input and output feature maps of the $i$-th convolutional layer, $Q_{\text{in}}$ and $Q_{\text{out}}$ are numbers of input and output neurons of FC layers.

\begin{table}[t]
	\caption{{Model Complexity Comparison}}
	\begin{center}
		\begin{tabular}{c|c|c|c}  
			\hline  
			{Model} & {Non-zero Parameters} & {FLOPs} & {HDF5} \\
			\hline  
			MAT-CENet & 1.03M & 37.01M  & 27.50M \\
			\hline
			CAT-CENet  & 0.70M & 28.93M & 28.70M \\
			\hline
			PCAT-CENet, $\kappa = 0.5$ & 0.64M & 28.93M & 9.63M \\
			\hline
			PCAT-CENet, $\kappa = 0.8$  & 0.61M & 28.93M & 9.63M \\
			\hline  
		\end{tabular}
	\end{center}
	\label{tab1}
\end{table}

Since FC and Conv2D layers in CAT-CENet are highly redundant, weight pruning is applied in these layers by sorting weights as per their absolute values and removing the smaller portion according to the pruning ratio $\kappa$, yielding PCAT-CENet. From Table I comparing the model complexity, the number of non-zero parameters and FLOPs of CAT-CENet are lower than MAT-CENet  \cite{S. Li}. PCAT-CENet further reduces the model size to facilitate the model storage while keeps the estimation accuracy almost unchanged, as shown in Fig.~\ref{Prune} in the following simulation results.

\subsection{Modality Contribution Analysis}

In this subsection, to justify the cross-modality gain of exploiting the radar sensing information for enhancing the communication channel estimation, the Shapley value, originating from game theory, is introduced. Specifically, the Shapley value is used to measure contributions of the communication branch input $\tilde{\mathbf{H}}_\mathrm{c}=[\tilde{\mathbf{H}}_\mathrm{c}^{\mathrm{R}},\tilde{\mathbf{H}}_\mathrm{c}^{\mathrm{I}}]$ and the sensing branch input $\tilde{\mathbf{A}}_k = [\tilde{\mathbf{A}}_k^{\mathrm{R}},\tilde{\mathbf{A}}_k^{\mathrm{I}}], k=1,\ldots,K$. With the output  $\hat{\mathbf{H}}_\mathrm{c}=[\hat{\mathbf{H}}_\mathrm{c}^{\mathrm{R}},\hat{\mathbf{H}}_\mathrm{c}^{\mathrm{I}}]$, the contribution of anyone of these input tensors is given by
\begin{eqnarray}
\label{eqn_shapley}
\varPhi = \sum_{i=1}^{M}\sum_{j=1}^{M} (\varphi_{i,j}^{\mathrm{R}}+\varphi_{i,j}^{\mathrm{I}}),
\end{eqnarray}
where $\varphi_{i,j}^{\ast}$ denotes the Shapley value of the $i$th element in $\hat{\mathbf{H}}_\mathrm{c}^{\ast}$ versus the $j$th element in $\tilde{\mathbf{H}}_\mathrm{c}^{\ast}$ or $\tilde{\mathbf{A}}_k^{\ast}$ with $\ast\in\{\mathrm{R}, \mathrm{I}\}$ representing the real or imaginary part. To reduce the computational complexity, $\varphi_{i,j}^{\ast}$ can be approximated as \cite{Shap}
\begin{eqnarray}
\label{eqn_apprx_shapley}
\varphi_{i,j}^{\ast} \approx \underset{\mathbf{B}_{0}\sim D, \atop \alpha\sim \mathcal{U}(0,1)}{\mathbb{E}} \left[(b_{j}^{\ast} - b_{0,j}^{\ast}) \frac{\partial f_{i}^{\ast}(\mathbf{B}_{0} + \alpha(\mathbf{B} - \mathbf{B}_{0}))}{\partial b_{j}^{\ast}} \right],
\end{eqnarray}
where $\mathbf{B}\in\{\tilde{\mathbf{H}}_\mathrm{c},\tilde{\mathbf{A}}_1,\ldots,\tilde{\mathbf{A}}_K\}$ and $\mathbf{B}_{0}$, randomly sampled from the dataset $D$, denote the original and reference input tensors, respectively, with $b_{j}^{\ast}$ and $b_{0,j}^{\ast}$ the corresponding $j$th elements. Besides, $f_{i}^{\ast}(\cdot)$ indicates the $i$th element in $\hat{\mathbf{H}}_\mathrm{c}^{\ast}$ and the scalar $\alpha$ uniformly ranging from $0$ to $1$ accounts for the change of the input from  $\mathbf{B}_{0}$ to $\mathbf{B}$. With the quantified contributions of the communication pilot and radar sensing information, $\varPhi$, the mechanism of the cross-modality gain can be revealed, as shown by Fig.~\ref{SHAP} in the following Section IV.

\section{Simulation Results}

In this section, simulation results are presented to demonstrate the superiority of the proposed CAT-CENet as well as to verify the cross-modality gain. The simulation parameters are configured as follows: the number of BS antennas $ M=256$, the wavelength $ \lambda=0.01$ meters, $\phi_l \sim U\left( -\frac{\pi}{2}, \frac{\pi}{2} \right) $, and $ r_l \sim U(10, 80)$ meters. The errors of the AoA and distance of sensing targets are modeled as $ \epsilon_{\theta_k^r} \sim \mathcal{C} \mathcal{N} \left({0},  {\sigma}_{\theta}^2 \right)$ with ${\sigma}_{\theta}=10^{-4}$ and $ \epsilon_{v_k} \sim \mathcal{C} \mathcal{N} \left({0}, {\sigma}_{v}^2 \right)$ with  ${\sigma}_{v}=10^{-2} $. The training, validation, and testing sets include 45,000, 5,000, and 2,000 samples, respectively. The normalized mean square error (NMSE), defined as $\mathbb{E}\{ \|{{\mathbf{H}}}_\mathrm{c} - \hat{{\mathbf{H}}}_\mathrm{c}\|_F^2 / \|{{\mathbf{H}}}_\mathrm{c}\|_F^2 \big\}$, is adopted to evaluate the estimation performance. The model is trained using the Adam optimizer with a batch size of 128 and an initial learning rate of $10^{-3}$ for 200 epochs. The baseline schemes for comparison include LS estimation, Linear Minimum Mean Square Error (LMMSE) estimation, Polar-domain Simultaneous Orthogonal Matching Pursuit (P-SOMP) \cite{M. Cui}, Mixed Attention Transformer based Channel Estimation Neural network (MAT-CENet) \cite{S. Li}, More-Information-Assisted Generative Adversarial Network (MIA-GAN) \cite{S. Luan}, and Score-Based Generative Models(SBGM) \cite{M. Arvinte}.
\begin{figure}[t]
\centering
\includegraphics[width=0.45\textwidth]{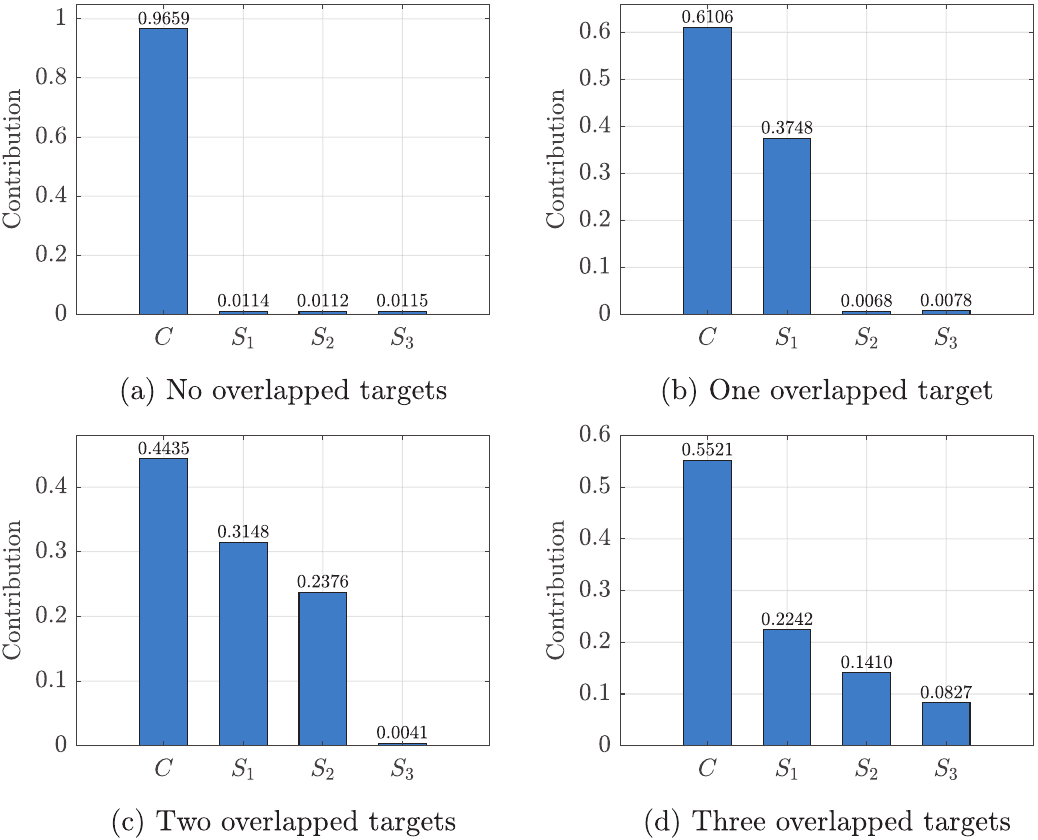}  
\caption{SHAP-based input contribution visualization under different numbers of overlapped targets.}
\label{SHAP}
\end{figure}

To verify the cross-modality gain of CAT-CENet, Fig.~\ref{SHAP} visualizes the contributions of input tensors with $L=K=3$ based on the Shapley additive explanation (SHAP) method. In SHAP, the Shapley value formulated in (\ref{eqn_apprx_shapley}) is used to measure the contribution. Specifically, the indices ``$C$", ``$S_1$", ``$S_2$", and ``$S_3$" of the horizontal axis in each subfigure denote the communication input tensor and the sensing input tensors corresponding to the first, second, and third targets, respectively.
The four subfigures display the change of the overlapping condition with the number of overlapped targets ranging from zero to three. In Fig.~\ref{SHAP}(a) with no overlapped targets, the communication pilot contributes almost all the information for the channel estimation while contributions of the three targets are negligible. When one target is overlapped as shown in Fig.~\ref{SHAP}(b), the contribution of this target increases significantly to $0.3748$ and becomes comparable with the communication input, leaving the contributions of other two non-overlapped targets still extremely small. This variation trend keeps for cases of two and three overlapped targets. This phenomenon demonstrates CAT-CENet can identify the features of overlapped targets and fully exploit them to enhance the communication channel estimation, as shown in the following results.

\begin{figure}[t]
\centering
\includegraphics[width=0.45\textwidth]{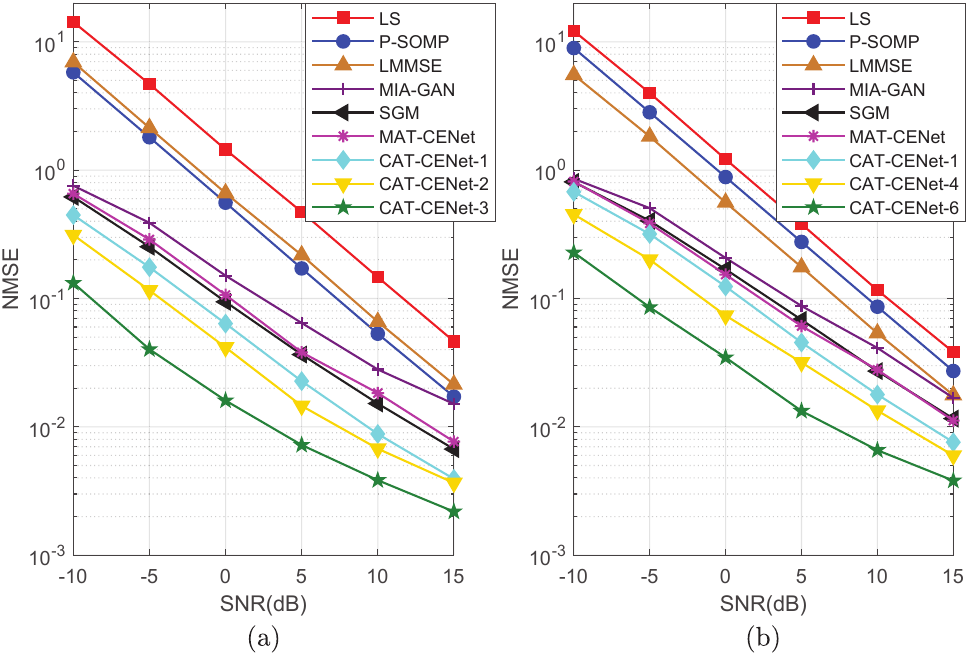}  
\caption{NMSE performance versus SNR of CAT-CENet and baseline schemes. (a) $L =K= 3$. (b) $L =K= 6$.}
\label{NMSE_comp}
\end{figure}

Fig.~\ref{NMSE_comp} compares the NMSE performance of CAT-CENet with baseline schemes versus signal-to-noise ratio (SNR) with $L =K= 3$ and $L =K= 6$, respectively, where ``CAT-CENet-1” indicates one overlapped targets and other legends follow the same notation. As shown in Fig.~\ref{NMSE_comp}, the proposed CAT-CENet consistently outperforms all baseline schemes under the whole SNR regime. In Fig.~\ref{NMSE_comp}(a), LS, LMMSE, and P-SOMP based on the traditional estimation theory achieve the ordinary performance. DL-based schemes outperform them significantly, especially in the low SNR. SGM and MAT-CENet achieve the evenly matched performance better than MIA-GAN. CAT-CENet achieves the best estimation performance and the advantage is remarkable even with only one overlapped targets. When SNR$=10$ dB, the NMSE of CAT-CENet-1 can approach to the order of magnitude of $10^{-3}$ while other schemes still maintain at $10^{-2}$. The performance gap will be enlarged if more overlapped targets are considered. The scenario with $L =K= 6$ follows the similar trends besides the overall performance degradation, revealing the negative impact of the increased number of paths on the estimation accuracy. However, it can be seen that CAT-CENet can compensate this performance loss by comparing CAT-CENet-1 in Fig.~\ref{NMSE_comp}(b) and SGM in Fig.~\ref{NMSE_comp}(a), which benefits from the elaborated structure of CAT-CENet enabling the better fusion of the heterogeneous communication and sensing features.

\begin{figure}[t]
	\centering
	\includegraphics[width=0.45\textwidth]{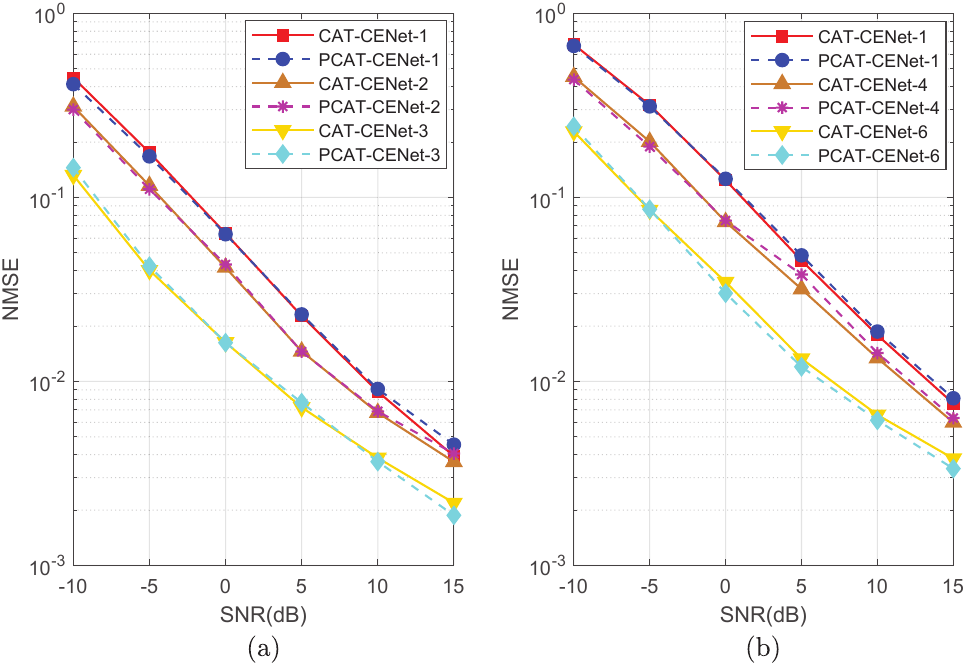}
	\caption{Performance comparison between PCAT-CENet and CAT-CENet.(a) $L =K= 3$. (b) $L =K= 6$.}
	\label{Prune}
\end{figure} 

To show the robustness of CAT-CENet to the weight pruning, Fig.~\ref{Prune} compares PCAT-CENet and CAT-CENet for $L =K= 3$ and $L =K= 6$ following the same notation as Fig.~\ref{NMSE_comp}, where the pruning ratio $\kappa = 0.8$. From Fig.~\ref{Prune}, PCAT-CENet achieves almost the same performance as CAT-CENet while can reduce the model size significantly. It also can be seen CAT-CENet in the scenario of $L =K= 3$ is more robust to the pruning, which can be even beneficial in the fully overlapped case.

\section{Conclusion}

In this article, CAT-CENet is proposed to enable the sensing-enhanced channel estimation for near-field ISAC XL-MIMO systems via the cross-modal fusion between the communication and sensing information, which is intractable for traditional methods and deep neural networks. The elaborated CAT-CENet can highlight and sufficiently exploit the information of those overlapped targets through the three-module structure and two-branch processing way. The Shapley value is introduced as the theoretical metric to quantify the input contribution and attempt to reveal the internal mechanism of the cross-modal gain. CAT-CENet can be simplified via the weight pruning with the almost unchanged accuracy. Simulation results show the superiority of CAT-CENet over the state-of-the-art schemes, which is more significant with the higher overlapping proportion.


\end{document}